\documentclass[amsmath,amssymb,aps,prd, preprintnumbers,nofootinbib,a4paper, 11pt]{revtex4-1}
\usepackage{amsthm}
\usepackage{color}
\usepackage{dcolumn}
\usepackage{bm}
\usepackage{bbm}
\usepackage{pxfonts}
\usepackage{slashed} 
\usepackage{url}

\usepackage[ margin=5pt, font=normalsize,labelfont=bf,justification=raggedright]{caption}

\usepackage[bookmarks, pagebackref=false]{hyperref}
\definecolor{rossoCP3}{cmyk}{0,.88,.77,.40}
\definecolor{verdeCP3}{rgb}{0.09765625, 0.57421875, 0.1015625}
\definecolor{bluCP3}{rgb}{0, 0.23, 0.67}
 
\newcommand{\ea}[1]{
\begin{align}
#1
\end{align}
}

\newcommand{\vevof}[1]{
\langle
#1
\rangle
}

\newcommand{\Tr}{\text{Tr}}

\newcommand{\be}{\begin{eqnarray}}
\newcommand{\ee}{\end{eqnarray}}

\begin{document}
\title{\Large  \color{rossoCP3} ~~  Conformal Extensions of the Standard Model \\ with \\ Veltman Conditions}
\author{Oleg Antipin}
\email{antipin@cp3-origins.net} 
\author{Matin Mojaza}
\email{mojaza@cp3-origins.net} 
\author{Francesco  Sannino}
\email{sannino@cp3-origins.net} 
\affiliation{{\color{rossoCP3} CP$^{3}$-Origins} \& Danish Institute for Advanced Study {\color{rossoCP3} DIAS}, University of Southern Denmark, Campusvej 55, DK-5230 Odense M, Denmark}

\begin{abstract}
 Using the renormalisation group framework we classify different extensions of the standard model according to their degree of naturality.  A new  relevant class of perturbative models involving elementary scalars is the one in which the theory simultaneously satisfies the Veltman conditions and is conformal at the classical level. We term these extensions perturbative natural conformal (PNC) theories. We show that PNC models are very constrained and thus highly predictive.  Among the several PNC examples that we exhibit, we discover a remarkably simple PNC extension of the standard model in which the Higgs is predicted to have the experimental value of  the mass equal to 126 GeV. This model also predicts the existence of one more standard model singlet scalar boson with a mass of 541 GeV and  the Higgs self-coupling to emerge radiatively. We study several other PNC examples that generally predict a somewhat smaller mass of the Higgs to the perturbative order we have investigated them.     
Our  results can be a useful guide when building extensions of the standard model featuring fundamental scalars.
\\[.1cm]
{\footnotesize  \it Preprint: CP$^3$-Origins-2013-039 DNRF90, DIAS-2013-39.}
 \end{abstract}

\maketitle

\thispagestyle{empty}
\tableofcontents
\newpage

\section{Degrees of (un)naturality}
  
With the discovery of the Higgs-like particle at CERN it has become imperative to critically investigate avenues that can lead to a deeper understanding of the phenomenon of mass generation in the standard model (SM), and that simultaneously are able to predict the scale of new physics.  

To be as general as possible we use the renormalisation group (RG) language to identify and classify the degree of naturality of the SM and its extensions. We start by analysing the renormalisation of the mass parameter of a simple $\phi^4$ real scalar field sector embedded in a more general theory. The Lagrangian terms we wish to discuss can be expressed  via the renormalised mass $m$, coupling $\lambda$ and the renormalised field $\phi_r$ \footnote{We impose a $Z_2$ symmetry which will be automatic when requiring the theory to be conformal at the classical level later.}:
\ea{
\mathcal{L}=\frac{1}{2}(\partial_{\mu}\phi_r)^2-\frac{1}{2}m^2 \phi_r^2-\frac{\lambda}{4!}\phi_r^4+\frac{\delta_Z}{2}(\partial_{\mu}\phi_r)^2-\frac{\delta_m}{2} \phi_r^2-\frac{\delta_{\lambda}}{4!}\phi_r^4 \ ,}
where the last three terms are counter terms needed to subtract the divergences plaguing the bare parameters ($\phi_B$, $m_0$, $\lambda_0$). The counter terms are defined as follows
\ea{
\phi_B\equiv \sqrt{Z}\phi_r \quad \delta_Z\equiv Z-1 \quad m^2\equiv m_0^2 Z-\delta_m \quad \delta_\lambda\equiv \lambda_0 Z^2 -\lambda \ .
}
The leading divergences to be accounted for by counter terms are $Z=1+f_1( \lambda, g_i)\log\frac{\Lambda^2}{m_0^2}+\dots$,
and $\delta_m= f_2(\lambda, g_i)\Lambda^2+\dots$, where $g_i$ denotes collectively the other renormalised dimensionless couplings of the theory. Here  $\Lambda$ is the cutoff of the theory. The explicit (e.g. leading order) expressions for the functions $f_1$ and $f_2$ in terms of the renormalised couplings $\lambda$ and $g_i$ are immaterial for the following discussion.  The only quadratically divergent parameter of the theory is the renormalised scalar mass which reads
\ea{
m^2=m_0^2(1+f_1(\lambda, g_i)\log\frac{\Lambda^2}{m_0^2})- f_2(\lambda, g_i)\Lambda^2 \ .
\label{massrenorm}
}
The expression above exemplifies the unnaturality of generic scalar field theories  at the quantum level. The problem being that even in the absence of an explicit mass term at the bare Lagrangian level a mass operator re-emerges via quantum corrections, living naturally at the highest energy scale of the theory $\Lambda$. Depending on where the cutoff energy scale is, compared to the physical value of the mass $m$ the couplings $\lambda$ and $g_i$ must be fine-tuned to achieve the value of $m$. Furthermore for the pure $\phi^4$-theory, $\lambda$ goes to zero when $\Lambda/m$ goes to infinity, i.e. the theory becomes non-interacting.

It is a valid question to ask whether one can consider partial or delayed solutions to the naturality problem above. We first categorise different proposals and then consider in detail a class of models that we define as {\it perturbative natural conformal} (PNC) extensions of the SM.

\subsection{Unnatural \& Classical conformality}
Unnatural models are extensions of the SM (or the SM itself) where the presence of any UV cutoff $\Lambda$ beyond the electroweak scale is dramatically felt because of the quadratic divergences of the theory.  Here as shown in Eq.~\eqref{massrenorm}, a large fine-tuning is required to keep the theory stable at the electroweak scale.  Constraints on these models will be of experimental nature, however some theoretical constraints can be enforced requiring, for example, the ground state of the theory to be stable up-to when the gravitational corrections are relevant \cite{Degrassi:2012ry,Antipin:2013sga}%
\footnote{ In \cite{Antipin:2013sga,Antipin:2013pya} it was shown that the quantum corrections need to satisfy the Weyl consistency conditions. According to Refs.~\cite{Degrassi:2012ry,Antipin:2013sga} the SM is in a metastable state and can therefore tunnel to the true ground state located at much higher values of the Higgs field.  The stability of the potential, per se, is lost at around $10^{10}$~GeV reinforcing the idea that one needs to go beyond the SM  to have a more complete theory of nature.}. 
 
Among the unnatural theories there is a growing popularity in the literature to investigate 
extensions of the SM, which are scale-invariant at the 
classical level of the action \cite{Foot:2007iy,Foot:2007as,Chang:2007ki,Meissner:2009gs,Meissner:2007xv,AlexanderNunneley:2010nw,Hur:2011sv,Englert:2013gz, Khoze:2013uia,Dermisek:2013pta,Farzinnia:2013pga,Carone:2013wla,Khoze:2013oga,Heikinheimo:2013fta}. In such extensions, the bare Higgs mass  $m_0$,
which is the only dimensionfull parameter of the SM, must be set to zero to enforce {\it classical} conformality. Of course, from Eq.~\eqref{massrenorm}, it is clear that unless the cutoff dependence is dropped there is no such thing as conformality in the theory. 
 Having dropped somewhat arbitrarily $\Lambda$, by resorting to e.g. dimensional regularisation, it is clear from Eq.~\eqref{massrenorm} that requiring $m_0=0$ at a given energy scale implies that $m=0$ at any RG scale. Electroweak symmetry can then be broken, for example, via the Coleman-Weinberg (CW)
mechanism\cite{Coleman:1973jx}. Here quantum corrections to the classical action induce the electroweak vacuum. 
The SM, on its own, cannot be classically conformal since the associated CW potential is unable  to lead to the observed Higgs mass \cite{Coleman:1973jx}. Therefore even classical conformality requires presence of new physics.

Notice, however, that the argument of resorting to dimensional regularisation to offset the cutoff is not a physical one. Conformality requires, as explained above, that all the directions in the parameter space of the theory to be (quasi) stable against quantum corrections. This means that unless there is no physical UV scale, quadratic divergences will be present. In other words one can set to zero the $\Lambda$ contribution only if the conformal model is free from any new scale; i.e. it is isolated. 
Once for instance gravity is taken into consideration it is hard to imagine how to avoid the reinstatement of the quadratic divergences.
This class of models is, however, interesting to investigate on pure theoretical grounds. For instance in Refs~\cite{Grinstein:2011dq, Antipin:2011aa,Antipin:2012kc,Antipin:2012sm} it was shown under controlled dynamics that these theories have a very rich structure and that a light scalar can emerge as a dilaton state at the quantum level.
These ideas have since found phenomenological applications in e.g. Refs.~\cite{Bellazzini:2012vz,Antipin:2013kia}.

\subsection{Delayed naturality}
Naturality of a scalar field theory can be delayed by requiring, when possible, that  the function $f_2$ in Eq.~\eqref{massrenorm} vanishes; i.e. the quadratic divergences are offset in perturbation theory as it was suggested long ago by Veltman \cite{Veltman:1980mj}. This simply postpones the energy scale at which the unnaturality of the theory re-emerges in RG time (See e.g. Ref.~\cite{Casas:2004gh,Masina:2013wja} for a recent analysis). Delayed naturality does not automatically imply or require conformality.  

It is well known that the SM again cannot delay its own naturality scale because, as we review later, the Veltman condition is not satisfied, and therefore the SM must be amended even in this case.

\subsection{Perturbative natural conformality (PNC)}
 Perturbative natural conformality (PNC) extensions of the SM combine classical conformality and delayed naturality. PNC has the beneficial effect to move the classically conformal extensions towards naturality while strongly restricting the parameter space of the four-dimensional extensions of the SM. This is the class of theories we wish to investigate here and for which we provide explicit examples.

\subsection{Natural}
A special status still beholds extensions which are free from quadratic divergencies such as electroweak scale Supersymmetry\cite{Feng:2013pwa} and Technicolor (see \cite{Sannino:2009za,Andersen:2011yj} for recent reviews). For supersymmetry the function $f_2$ vanishes to all orders, while for Technicolor models the cutoff of the theory is identified with the scale of compositeness. 

\section{PNC conditions}
It is useful to set the stage by reviewing the Coleman-E. Weinberg (CW) phenomenon of spontaneous symmetry breaking \cite{Coleman:1973jx} following the more general analysis of Gildener and S. Weinberg \cite{Gildener:1976ih}. 

Consider a theory with a number of weakly-coupled real scalar fields $\phi_i$, with $i$ counting the scalars. It is convenient to renormalise their masses at the origin of the scalar-field space to be zero.  
The renormalised, scale-invariant, tree-level scalar potential then reads:
\be
\label{treelevelpotential}
V_0(\phi_i)=\frac{\lambda_{ijkl}}{24}\phi_i\phi_j\phi_k\phi_l+ \text{fermionic and vectorial contributions}+ c.t.
\ee
where $c.t.$ stands for counter terms and $\lambda_{ijkl}$ are renormalized quartic couplings. 
The scalar fields are renormalised at the scale $\mu_0$ such that here they are elementary; i.e. have zero anomalous dimensions.
In general the theory will also contain gauge and Yukawa couplings. 
We denote these globally by $g$ and $y$, respectively. 
Consistent perturbation theory requires
\ea{
\label{loopcondition}
\lambda_{ijkl} \sim g^2 \sim y^2 \ll 1 \ ,
}
for any non-zero $\lambda_{ijkl}$.
Thus 
the effective potential is, in general, dominated by the tree-level potential. 
%
The way in which loop corrections can
shift the global minimum to a non-zero point in scalar field space, is if the tree-level potential
has flat directions in field space, $n_i$, 
within the perturbative regime of Eq.~\eqref{loopcondition}.
The renormalisation condition (i.e. here the starting point of perturbation theory) we must therefore impose on the set of quartic couplings
to make a viable CW analysis,
is to constrain the parameter space of the couplings to a subspace,
where the renormalised tree-level potential does have flat directions.
This fixes the renormalisation scale $\mu$ to a specific value $\mu_0$, 
given by the renormalisation condition on the quartic couplings.
Taking $u_i$ to be a unit vector in field space, the flat directions
can be found by solving the problem:
\ea{
{\rm min}(\lambda_{ijkl} u_i u_j u_k u_l ) \big |_{u_i u_i = 1} = 0 \ .
}
If a solution $u_i = n_i$ exists, then $\phi_i = n_i \phi$ is a flat direction 
of the tree-level potential, along which the CW analysis can
consistently be made. Thus the renormalisation condition on the quartic
couplings read:
\ea{
\label{renormalisationcondition1}
\lambda_{ijkl} (\mu_0) n_i n_j n_k n_l \approx 0 \ . 
}
By using $\approx$ it is implied that the condition has to be satisfied at least
to the order $g^4$ in the quartic couplings, that is, the renormalisation condition
can be relaxed to the level:
\ea{
\label{renormalisationcondition2}
\lambda_{ijkl} (\mu_0) n_i n_j n_k n_l \sim \mathcal{O}(g^4) \ . 
}
Anything beyond this is not a viable setup to study the CW mechanism\footnote{In the original paper, Coleman and Weinberg first used
scalar electrodynamics in which there is only one quartic coupling, $\lambda$.
The renormalisation condition they imposed was $\lambda(\mu_0) \sim \mathcal{O}(e^4)$ where $e$ is the electric charge, consistent with Eqs.~\eqref{renormalisationcondition1} and \eqref{renormalisationcondition2}.
In the case of multiple quartic couplings the renormalisation condition Eq.~\eqref{renormalisationcondition1} should first be used (together with \eqref{loopcondition}). This defines the viable parameter space for the CW analysis, and then perturbations
of the order $g^4$ from this hypersurface can be investigated.}.

Now, consider the one-loop correction to the effective potential in the Landau gauge on some classical background $\phi_c = u_i \phi_i$, with $u_i$ a unit vector:
\be
\label{oneloopterm0}
V_1(\phi_c)&=&\frac{1}{2}\int \frac{d^4 k}{(2\pi)^4} \ Str \big[ \ln (k^2+M^2(\phi_c))\big]+c.t. 
\ee
where 
$M^2(\phi_c)$ is the background dependent mass matrix and we defined the supertrace
\ea{
\label{Str}
Str\equiv \sum\limits_{scalars}-\ 2\sum\limits_{Weyl fermions}+\ 3\sum\limits_{vectors} \ .
}
We consider the theory to be a low-energy description valid up to some scale $\Lambda$ and regularize the integral in the UV with a hard cutoff at $k = \Lambda$. Furthermore one must assume $\phi_c \neq 0$ such that the integral does not diverge in the IR. The one-loop contribution is then straightforwardly computed and after an expansion in $M^2(\phi_c) \ll \Lambda^2$ we find:
\ea{
\label{oneloopterm}
V_1(\phi_c) = \frac{1}{64 \pi^2} Str\left [ \Lambda^4 \left ( \ln \Lambda^2- \frac{1}{2} \right ) + 2 M^2(\phi_c) \Lambda^2+ M^4(\phi_c) \left ( \ln \frac{M^2(\phi_c)}{\Lambda^2}- \frac{1}{2} \right ) \right ] + c.t.
}
The first term is the cosmological constant term, which vanishes in a theory with equal number of bosonic and fermionic degrees of freedom. In this work we will have nothing more to say about the cosmological constant term and therefore subtract it away from the potential.

The scalar masses are computed from the full potential via $m_i^2 = \partial^2 V/\partial \phi_i^2$. It is the second term that is responsible for the quadratic divergence of the scalar masses.
These divergences can either be removed by appropriate choices for the counter-terms (fine-tuning), making the theory unnatural, or vanish identically if $\partial^2Str M^2(\phi_i)/\partial \phi_i^2=0$ to all orders due to e.g. symmetry reasons (such as supersymmetry) and are thus natural theories.

Naturality can be partially achieved, or better said delayed, by imposing the Veltman conditions defined such that, to the same perturbative order used by the CW analysis, the quadratic divergences appearing from Eq.~\eqref{oneloopterm} in the mass of any non-Goldstone scalar must vanish:
\be
\frac{1}{2}\frac{\partial^2 Str[M^2(\phi_i)]}{\partial \phi_i^2} \ \bigg|_{\mu_0}=0 \ .
\label{Veltman}
\ee
This leads to extra constraints on the dimensionless couplings of the theory.  

Considering now the CW analysis along any tree-level flat direction $\phi = n_i \phi_i$,
together with the Veltman conditions, at a chosen renormalisation scale $\mu_0$ yields the following one-loop effective potential 
\ea{
\label{AB}
V(\phi_c)=V_1(\phi_c)= \frac{1}{64\pi^2} Str \left[M^4(\phi_c) \ln \frac{M^2(\phi_c)}{\mu_0^2} \right]  = A \phi_c^4 + B \phi_c^4 \ln \frac{\phi_c^2}{\mu_0^2}\ ,
}
where we set the counter terms to properly renormalise the couplings
\ea{
\frac{\partial^4 V}{\partial \phi_c^4} \Bigg |_{M^2(\phi_c) = \mu^2 e^{-25/6}} = \lambda_{ijkl}(\mu) n_i n_j n_k n_l \ ,
} 
and used that $ \lambda_{ijkl}(\mu_0) n_i n_j n_k n_l =0$. 
The scaling factor $e^{-25/6}$ is chosen only to simplify Eq.~\eqref{AB} and corresponds
to the renormalization choice made in \cite{Gildener:1976ih}.

In a classically scale-invariant theory, all masses will be proportional to $\phi_c$ and thus we can write $M^2(\phi_c) = W^2 \phi_c^2$, such that in Eq.~\eqref{AB}:
\ea{
A = \frac{1}{64\pi^2} Str W^4 \ln W^2 \ , \quad 
B= \frac{1}{64\pi^2} Str W^4 \ .
} 
The non-trivial stationary point of the effective potential is at:
\be
\log\frac{\langle \phi_c^2 \rangle}{\mu_0^2}=-\frac{1}{2}-\frac{A}{B} \ ,
\ee
and since both functions $A$ and $B$ appear at one loop we have that $\phi_c\sim \mu_0$ and therefore perturbation theory is valid, as expected by construction.
If the extremum corresponds to the ground state of the theory we have for the  scalar fluctuation, along the classical flat direction, a positive mass squared which reads
\be
\label{oneloopmass}
m^2_{CW}= 8 B \langle \phi_c^2 \rangle \ . 
\ee
 The masses of the other non-Goldstone scalars arise at tree-level and are  positive as well \cite{Gildener:1976ih}.
 
 \section{PNC Models }
 \noindent
Having set up the stage for PNC we move on to examine specific models.

\subsection{Standard Model}
In the SM the renormalised tree-level potential, including the gauge and Yukawa terms, reads
\ea{
V_0^{SM} = \lambda \left (H^\dagger H \right)^2  - \frac{1}{2}Ê\left (g^2 W_\mu^+ W^{ - \mu} + \frac{g^2+{g^\prime}^2}{2} Z_\mu Z^\mu \right ) H^\dagger H + y_t (\bar{t}_L , 0) \left(i\sigma^2 H^*\right) t_R + {\rm h.c.} +  \
c.t. \ ,
}
where we have set the renormalised mass to zero and neglected the Yukawa couplings to the leptons and light quarks (with respect to the top-quark).

To compute the Veltman condition for the Higgs mass, we expand the Higgs doublet around the electroweak background: $H = \frac{1}{\sqrt{2}} ( \pi_2 +i \pi_1, v+ h - i \pi_3) $, 
and compute the mass-matrix, keeping only the $h$-dependent part, which is what will remain in the Veltman condition for $h$:
\be
\label{SMmassmatrix}
\frac{M^2(h)}{h^2} =\text{diag} \left \{ 3\lambda ,\ \lambda, \ \lambda, \ \lambda, \ \frac{1}{4}g^2, \ \frac{1}{4}g^2, \ \frac{1}{4}(g^2+g^{\prime 2}), \ \frac{1}{2}y_t^2, \ \frac{1}{2}y_t^2 \right\} \ ,
\ee
where the entries correspond respectively to the mass of the Higgs boson, the three (would be) Goldstone bosons, the $W^+$, $W^-$ and $Z$ vector bosons and two top quark color multiplets in the Weyl basis.
Then from Eq.~\eqref{Veltman} follows Veltman's condition for the Higgs mass:
\be
\label{VeltmanSM}
\frac{1}{2}\frac{\partial^2 Str[M^2(h)]}{\partial h^2} \ \bigg|_{\mu_0}= {6}\lambda(\mu_0)+\frac{9}{4}g^2(\mu_0)+\frac{3}{4}g^{\prime 2}(\mu_0)-6y_t^2(\mu_0)=0 \ .
\ee
Note that this condition is independent of the vev and that once the vacuum is generated, the Veltman conditions for the Goldstone directions disappear.

To generate the vev through the CW mechanism, we must
assume the tree-level potential to be flat at the same scale $\mu_0$ at which the Veltman condition is imposed:
\be
 \lambda(\mu_0) \approx 0 \ .
\label{SMflat}
\ee
The Veltman condition under this constraint reduces to:
\ea{
3g^2(\mu_0)+g^{\prime 2}(\mu_0)-8y_t^2(\mu_0)=0 \ .
}
Requiring this relation to hold, while using $\mu_0 \sim v \approx 246$~GeV and keeping $m_W^2 = v^2 g^2(\mu_0)/4$ and  \mbox{$m_Z^2 = v^2(g^2(\mu_0)+{g^\prime}^2(\mu_0))/4$} at their physical value, leads to a too light top quark mass \cite{Chaichian:1995ef}:
\ea{
4m_t^2=m_Z^2+2m_W^2 \quad \Longrightarrow \quad  m_t\approx  73 \ \text{GeV} \ .
}
The Higgs mass is induced at one-loop, which is given by \eqref{oneloopmass} and reads
\ea{
m^2_{h} &= \frac{3}{8\pi^2} \big[\frac{1}{16}\big(3g^4+2 g^2 g'^2+g'^4\big)+  {4} \lambda^2-y_t^4\big] v^2 \nonumber \\
& \stackrel{\mu=\mu_0}{=} \frac{3}{512 \pi^2} \left ( 3 g^4 + 2 g^2 {g^\prime}^2 + 3{g^\prime}^4 \right ) v^2  \quad \Longrightarrow \quad  m_h\approx  5 \ \text{GeV}\ .
}
This example shows that the PNC conditions are quite constraining. 
In fact, as it is well-known, working with only one of the conditions, either Veltman's condition or the CW condition, one would in the first case find a too large Higgs mass and the second case a too low Higgs mass. The example also shows the predictive power of a PNC-like model, which here predicts (wrongly) both the top and Higgs mass.

\subsection{SM + singlet scalar}\label{singletscalar}
We next consider the simplest conformal extension of the SM, where a {\it real} scalar singlet $S$ is added.  The CW phenomenon in this model has been studied in Ref.~\cite{Chang:2007ki, Foot:2007as}. Here, we review and extend the analysis by considering the additional constraints imposed by the Veltman condition.
The requirement of classical conformality together with renormalisability leads to the following $Z_2$ symmetric potential 
\ea{
\label{SMsingletpot}
V_0 = V_0^{SM} + \lambda_{HS} H^\dagger H S^2 + \frac{\lambda_S}{4} S^4 + c.t. 
}
The constraint from requiring the potential to be bounded from below is found by completing the square and reads:
\ea{
\label{bounded}\lambda \geq 0 \ , \quad \lambda_S \geq 0 \ ,  \quad  \text{and if \ $\lambda_{HS}<0$ :}\quad \lambda \lambda_{S} \geq \lambda_{HS}^2  \ .
}

Before proceeding to the one-loop CW analysis, we impose the Veltman conditions on the couplings to cancel the quadratic divergences at one loop.
The Veltman condition for $S$ is simple to compute and reads:
\be
\label{VeltmanS1}\frac{1}{2}\frac{\partial^2 Str[M^2(S)]}{\partial S^2} \ \bigg|_{\mu_0}=3\lambda_S(\mu_0)+4\lambda_{HS}(\mu_0) =0 \ .
\ee
We observe immediately that this condition can only be satisfied if $\lambda_{HS} < 0$.
The Veltman condition for the Higgs doublet is derived as described in the previous section. The mass matrix in Eq.~\eqref{SMmassmatrix} now has an additional entry from the field $S$, which is simply $\lambda_{HS}$. The Veltman condition for the Higgs field $h$ thus reads:
\be
\label{VeltmanSinglet}
\frac{1}{2}\frac{\partial^2 Str[M^2(h)]}{\partial h^2} \ \bigg|_{\mu_0}= {6}\lambda(\mu_0)+\frac{9}{4}g^2(\mu_0)+\frac{3}{4}g^{\prime 2}(\mu_0)-6y_t^2(\mu_0)+ \lambda_{HS}(\mu_0)=0 \ .
\ee
We note that there are no further one loop Veltman conditions once the electroweak vacuum is generated, since the remaining three scalar degrees of freedom will turn into Goldstone fields. Moreover, in the $H$ and $S$ basis there are no off-diagonal quadratic divergences at one loop.

We next consider the one-loop CW analysis under the above two constraints as a mechanism to generate the electroweak vacuum radiatively.
To study the possible classical moduli of the scalar potential, it is sufficient to consider the Higgs doublet in a unitary gauge, where it reduces to one degree of freedom.
It is then useful to reparametrize the scalar fields in terms of polar coordinates:
\ea{
H = \frac{r}{\sqrt{2}} \begin{pmatrix} 0 \\ \cos \omega \end{pmatrix} \ , \quad S = r \sin \omega \ ,
}
such that the tree-level potential of the scalar sector simplifies to:
\ea{
V_0 = \frac{r^4}{4}\left( \lambda \cos ^4\omega+\lambda _{S}\sin ^4\omega
   +2 \lambda _{HS} \sin ^2\omega \cos ^2 \omega \right) +
  c.t.
}
The minima of this potential will be along the ray $r$ in some unit direction $n = (\cos \vevof{ \omega} , \sin \vevof{\omega})$. These are found by studying the first and second derivatives of the tree-level potential.
The results are:
\ea{
0\leq\lambda < \text{min}\{\lambda_S, \lambda_{HS}\} \ &: \quad \vevof{\omega} = 0 \ , \label{constraint}\\
0\leq \lambda_{S} < \text{min}\{\lambda, \lambda_{HS}\} \ &: \quad \vevof{\omega} = \frac{\pi}{2} \ , \label{constraint2}\\
- \sqrt{\lambda \lambda_S} \leq \lambda_{HS} < \text{min}\{\lambda, \lambda_S\} \ & : \quad  
\tan^2\vevof{\omega} = \frac{\lambda-\lambda_{HS}}{\lambda_S - \lambda_{HS}} \ .
\label{nontrivialvev}
}
Considering the CW analysis one can, for the cases $\lambda_{HS}> \text{max}\{\lambda,\lambda_S\}$, also study metastable flat directions along either $\omega =\pi/2 $ or $\omega =0$.
{}For $\vevof{\omega} = 0$ it is clear that only $h$ gets a vev and the CW analysis follows the SM case. The $\vevof{\omega} = \pi/2$ case is similar to that analysis, but does not lead to electroweak symmetry breaking.
In neither case, however, the Veltman conditions can be satisfied, since Eq.~\eqref{VeltmanS1} requires $\lambda_{HS}<0$. 
We conclude that in these two cases we cannot satisfy the PNC conditions.

We analyse now to the third possibility, i.e. Eq.~\eqref{nontrivialvev} 
to investigate whether the PNC requirements can be satisfied.
First, we find the renormalisation condition from Eq.~\eqref{renormalisationcondition1}, which sets the tree-level potential to zero along the $\vevof{\omega}$-direction.
It is given by $\lambda \lambda_S - \lambda_{HS}^2 = 0$ and can also be expressed as:
\ea{
\left(\sqrt{\lambda (\mu_0) \lambda_S(\mu_0)} - \lambda_{HS}(\mu_0)\right)
\left(\sqrt{\lambda (\mu_0) \lambda_S(\mu_0)} + \lambda_{HS}(\mu_0)\right) = 0+ \mathcal{O}(\lambda^4) \ .
}
From the Veltman condition Eq.~\eqref{VeltmanS1} we need $\lambda_{HS} < 0$,
and thus setting the first parenthesis to zero is not viable.
We must therefore require the second parenthesis to vanish at the renormalisation scale $\mu_0$:
\ea{
\sqrt{\lambda (\mu_0) \lambda_S(\mu_0)} + \lambda_{HS}(\mu_0) = 0 + \mathcal{O}(\lambda^2) \ .
\label{scalar-flat}}
This relation saturates the stability bound of the potential, given in Eq.~\eqref{bounded} and the tree-level potential at the scale $\mu_0$ simplifies to:
\ea{
V_0(\mu_0) = \lambda \left ( H^\dagger H - \frac{|\lambda_{HS}|}{2 \lambda} S^2\right ) ^2  +  c.t.
}
Thus the PNC requirement has lead  to an $SO(4,1)$ symmetric tree-level potential for the scalar sector.

Now, it follows from the CW analysis that the one-loop contribution along the tree-level flat direction $\vevof{\omega}$ will give a non-trivial vev at some value $\vevof{r}$, 
for parameter values that gives a positive curvature. 
The electroweak vev fixes the value of $\vevof{r}$ through
\ea{
\vevof{r} \cos \vevof{\omega} = v  \approx 246 \text{ GeV} \ , \qquad {\rm and~we~take } \qquad {v \approx \mu_0} \ .
}
Rewriting $r \cos \vevof{\omega}$  as $(v + h)$ and $r \sin \vevof{\omega}$ as $v \tan \vevof{\omega} + s$,
we parametrize the light and heavy mass eigenstates by:
\ea{
\phi = h \cos\vevof{\omega} + s \sin \vevof{\omega} \ , \quad 
\Phi =s \cos\vevof{\omega} - h \sin \vevof{\omega}  \ ,
}
which have the tree-level masses:
\ea{
m_{0,\phi}^2 = 0 \ , \quad m_{0,\Phi}^2 = 2 (\lambda-\lambda_{HS}) v^2 \ .
}

The mass of $\phi$ emerges at one loop, since it is the field along the tree-level flat direction. 
Its one-loop mass is given by Eq.~\eqref{oneloopmass} and reads:
\ea{
m_{1,\phi}^2 &= \frac{1}{8\pi^2} \frac{Str M(\vevof{r})^4}{\vevof{r}^4} \vevof{r}^2
= \frac{\cos^2\vevof{\omega}}{8\pi^2v^2} [ 6 m_W^4 + 3 m_Z^4 + m_\Phi^4 - 12 m_t^4] \nonumber \\
&= \cos^2\vevof{\omega}\frac{v^2}{8\pi^2} 
\left [\frac{6}{16} g^4 + \frac{3}{16}(g^2+{g^\prime}^2)^2+ 4(\lambda-\lambda_{HS})^2 -  \frac{12}{4} y_t^4 \right] \ .
}

Imposing now the Veltman conditions at the scale $\mu_0$, where $\lambda_{HS}^2 = \lambda \lambda_S$, we get that
\ea{
&\lambda_{HS}(\mu_0) = - \frac{3}{4} \lambda_S(\mu_0) \ ,  \ \lambda_S(\mu_0) = \frac{16}{9} \lambda(\mu_0) \ ,  \ \cos^2\vevof{\omega} = \frac{4}{7} \ , \\
&\lambda(\mu_0) = \frac{9}{56} \left ( 8 y_t^2(\mu_0) -3 g^2(\mu_0) - {g^\prime}^2(\mu_0) \right) \ .
}
Thus, all parameters of the model are fixed from the experimental values of the top-quark mass and the $W$ and $Z$ boson masses.
The renormalization scale is approximately \mbox{$\mu_0 \approx v = 246$ GeV}. Using the experimental values for the mass of the top, W and Z at this scale, we get: 
\ea{
m_\phi \approx 95 \text{ GeV} \ , \quad m_\Phi \approx 541 \text{ GeV} \ .
}
The state $\phi$ is to be identified with the Higgs boson and
is a mixed state of $h$ and $S$ with mixing angle $\vevof{\omega} \approx 0.2 \pi$, making it mostly $h$-like.
This result implies that the PNC extensions of the SM with just one real scalar can lead to spectra close to the observed particle masses. In addition it requires the existence of yet another heavier scalar. It would be interesting to go beyond the one-loop analysis to investigate whether one can recover the observed value of the Higgs mass. One should simultaneously also investigate the effects of the mixing angle which will partially modify the Higgs phenomenology. 

This example also shows that quadratic divergences can cancel in a purely scalar sector, due to an approximate $SO(4,1)$ symmetry. 
%

\subsection{$S\bar{E}\chi y$ extension of the SM }\label{Sexy}
We investigate next a model containing, besides a complex scalar, also two Dirac fermions.  The model was introduced in Ref.~\cite{Dissauer:2012xa} to provide an explicit calculable example of a magnetic dark matter  extension of the SM.
It consists of a vector-like heavy electron, $E$, a {complex} scalar electron, 
$S$
, and a SM singlet Dirac fermion, $\chi$, playing the role of the dark matter. 
The tree-level potential of the model, including Yukawa interactions, reads:
\begin{equation}
V_0 =V_0^{SM}+\lambda_S(S^\dagger S)^2+\lambda_{HS}H^\dagger H S^\dagger S+(S\bar{E}\chi y +{\rm h.c.}) + { c.t.}  
\end{equation}
The interactions among the $\chi$ and the SM particles occur via loop-induced processes involving the Yukawa coupling $y$. The heavy electron $E$ and the scalar electron $S$ both carry SM hypercharge which is equal to their electric charge $Q_E=Q_S=1$. This is a microscopic realization of the effective model introduced in \cite{DelNobile:2012tx}  to accommodate several direct dark matter search experiments.

We require the potential to be bounded from below which leads
to the constraints: $\lambda \geq 0$, $\lambda_S \geq 0$ and if $\lambda_{HS}<0$ then $\lambda_{HS}^2 \leq 4 \lambda \lambda_S$,
which is a modification of Eq.~\eqref{bounded}, due to a different normalization of $\lambda_{HS}$.
We further demand that $\langle S\rangle=0\ $, so that we do not break the $U(1)$ symmetry of electromagnetism. This implies to use Eq.~\eqref{constraint}: $0\leq \lambda < \text{min}\{\lambda_S, \lambda_{HS}/2\}$. Furthermore, to have a flat direction in the Higgs potential we require Eq.~\eqref{SMflat}, i.e. that $ \lambda(\mu_0)=0$. 
The tree-level mass matrix on the $h$-background reads:
\be
\frac{M^2(h)}{h^2} =\text{diag}(3\lambda ,\ \lambda,\ \lambda,\ \lambda,\ \frac{1}{4}g^2, \ \frac{1}{4}(g^2+g^{\prime 2}), \ \frac{1}{2}y_t^2, \ \frac{1}{2}y_t^2, \ \frac{1}{2}\lambda_{HS},   \ \frac{1}{2}\lambda_{HS}) \ ,
\ee
where the last two entries are the tree-level masses of the real components of the scalar electron and the first part is equivalent to the SM case Eq.~\eqref{SMmassmatrix}.
The Veltman condition for the Higgs mass reads:
\be
\label{bound1}\frac{1}{2}\frac{\partial^2 Str[M^2(h)]}{\partial h^2} \ \bigg|_{\mu_0}=\frac{9}{4}g^2(\mu_0)+\frac{3}{4}g^{\prime 2}(\mu_0)-6y_t^2(\mu_0)+\lambda_{HS}(\mu_0) =0 \ ,
\ee
while for $S$ it reads:
\be
\label{VeltmanS} \frac{1}{2}\frac{\partial^2 Str[M^2(S)]}{\partial S^\dagger \partial S} \ \bigg|_{\mu_0}=4\lambda_S(\mu_0)+2\lambda_{HS}(\mu_0) + {3}{g^\prime}^2(\mu_0)-2y^2(\mu_0)=0 \ .
\ee
Here the  Veltman condition for $S$ can be satisfied due to the $S\bar{E}\chi$y operator.
There is a unique family of solutions to the Veltman conditions 
under the flatness constraint $\lambda(\mu_0) \approx 0$:
\ea{
\lambda_{HS}(\mu_0) &= \frac{3}{4} \left( 8y_t^2(\mu_0) - 3 g^2(\mu_0) - {g^\prime}^2(\mu_0) \right ) \stackrel{\mu_0 \approx v}{ \approx} 4.84 \ , \\
\lambda_{S}(\mu_0) &= \frac{y^2(\mu_0)}{2} - \frac{1}{2}\lambda_{HS}(\mu_0) - \frac{3}{4} {g^\prime}^2(\mu_0)  \stackrel{\mu_0 \approx v}{ \approx} \frac{y^2(\mu_0)}{2} - 2.77 \ ,
}
where in the last equalities we used the experimental values for the couplings. Due to the stability bound we have $y(\mu_0) \geq 2.35$.

From these constraints we arrive at a prediction for the mass of $S$ and the one-loop mass of the Higgs:
\ea{
\label{HiggsCW}
m^2_{h} = \frac{3}{8\pi^2} \big[\frac{1}{16}\big(3g^4+2 g^2 g'^2+g'^4\big)+\frac{1}{6}\lambda_{HS}^2-y_t^4\big] v^2 
\quad &\Longrightarrow \quad m_h \approx 83 \text{ GeV} \ , \\
m_S^2 = \frac{1}{2} \lambda_{HS} v^2 \quad &\Longrightarrow \quad m_S \approx 383 \text{ GeV} \ .
}
The mass of $S$ is within LHC reach and coincidently it has about the same value used as benchmark in \cite{Dissauer:2012xa}. The Higgs is lighter than the experimentally observed one.  However due to the relatively large value of $\lambda_{HS}$, higher order corrections can be relevant.  
The phenomenological consequences of the model without requiring conformality but imposing the Veltman conditions are being investigated in \cite{Frandsen:2013bfa}. 

\section{ Concluding with an intriguing PNC candidate }
From the above it is clear that the PNC models are quite constrained and therefore highly predictive. 
We conclude by presenting an intriguing model where, at the one-loop level, one finds an Higgs with the observed value of the mass, while predicting yet another massive scalar around $540$~GeV.   The model is surprisingly simple, consisting of just another real scalar $S$ and a Weyl fermion, $\chi$. The potential of the theory, together with the Yukawa interaction between $S$ and $\chi$, is
\ea{
V_0 = V_0^{SM}+ \lambda_{HS} H^\dagger H S^2 + \frac{\lambda_S}{4} S^4 +y_\chi S (\chi \chi +   \bar{\chi} \bar{ \chi}) +  c.t. 
}
Here we are using the Wess-Bagger notation for the Weyl fermion. 
The scalar sector is the same as in Eq.~\eqref{SMsingletpot} and the stability bound is therefore given by Eq.~\eqref{bounded}.
We study the case  where the vev of $S$ vanishes which implies Eq.~\eqref{constraint} to hold. 
The Veltman conditions read
\ea{
\frac{1}{2}\frac{\partial^2 Str[M^2(S)]}{\partial S^2} \ \bigg|_{\mu_0}=3\lambda_S(\mu_0)+4\lambda_{HS}(\mu_0)-8y_\chi^2 =0 \ ,
}
and
\be
\frac{1}{2}\frac{\partial^2 Str[M^2(h)]}{\partial h^2} \ \bigg|_{\mu_0}= 
6 \lambda(\mu_0) + \frac{9}{4}g^2(\mu_0)+\frac{3}{4}g^{\prime 2}(\mu_0)-6y_t^2(\mu_0)+\lambda_{HS}(\mu_0)=0 \ .
\ee
The first condition can now be satisfied due to the presence of the Yukawa coupling $y_\chi$ while the second condition is identical to Eq.~\eqref{VeltmanSinglet}.
For the CW analysis to work, we impose $\lambda(\mu_0) \approx 0$
and thus the solution to the Veltman conditions are:
\ea{
\lambda_{HS}(\mu_0) &= 6y_t^2(\mu_0)-\frac{9}{4}g^2(\mu_0)-\frac{3}{4}g^{\prime 2}(\mu_0) \stackrel{\mu_0 \approx v}{\approx} 4.84 \ , \\
\lambda_S(\mu_0) &= \frac{8}{3} y_\chi^2(\mu_0)-\frac{4}{3} \lambda_{HS}(\mu_0) \stackrel{\mu_0 \approx v}{\approx} \frac{8}{3} y_\chi^2(\mu_0) - 6.45 \ ,
}
where we have used \mbox{$\mu_0 \approx v = 246$ GeV} and the experimental values for the masses of the top quark and the W and Z bosons.
The second solution sets a lower bound on $y_\chi$ from the stability bound on $\lambda_S$, i.e. $y_\chi(\mu_0) \geq 1.55$.

From these constraints we arrive at a prediction for the one-loop induced Higgs mass, and for the tree-level mass of $S$:
\ea{
m^2_{h} &= \frac{3}{8\pi^2} \big[\frac{1}{16}\big(3g^4+2 g^2 g'^2+g'^4\big)-y_t^4+ \frac{\lambda_{HS}^2}{3}\big] v^2  \quad  
&\Longrightarrow \quad m_h \approx 126 \ \text{GeV}\ , \\
m_S^2 &= \lambda_{HS} v^2 \quad 
&\Longrightarrow \quad m_S \approx 541 \ \text{GeV}\ .
}
These PNC  predictions do not depend on the specific details of the extra fermionic sector.  Given the relatively large values of the couplings, albeit still in the perturbative regime, it is relevant to investigate the higher order corrections. 
\\ 

To summarize, we classified the degree of naturality of SM extensions using the renormalisation group framework, and introduced the concept of perturbative natural conformality (PNC).
To further appreciate the relevance of the PNC conditions we provide, in the Appendix, one last example featuring a Gauge-Yukawa theory possessing IR fixed points.

We have shown that the PNC framework can be highly predictive and can lead to realistic extensions of the SM. In particular PNC models have the generic feature to predict new states within LHC reach. Another generic feature of these models is that the Higgs self-coupling differs from the SM one.

\acknowledgements
We thank Marc Gillioz, Jens Krog, Esben M\o lgaard and Ole Svendsen for useful discussions and comments.
This work was supported by the Danish National Research Foundation DNRF:90
grant.
 
\newpage
\appendix

\section{Gauge-Yukawa theories with fixed points and PNC conditions}

In the main text we focused on PNC extensions of the SM which did not feature any IR or UV fixed points. However, theories with fundamental scalars can possess, both infrared and ultraviolet fixed points. Because of the presence of fundamental scalars, and in absence of supersymmetry, these fixed points are all unstable with respect to the addition of a scalar mass operator, at least within perturbation theory. It is therefore valuable to soften the effects of an UV cutoff by requiring the models to abide the PNC conditions. 
We will show here that imposing the PNC conditions leads to relevant constraints for the phase-diagram of the theory.  Our analysis can be extended to any scalar field theory featuring fixed points.

 We use the model studied in \cite{Antipin:2011aa} in which we investigated the infrared dynamics of a non-supersymmetric SU(X) gauge theory featuring an adjoint fermion, $\lambda_m$, $N_f$ Dirac flavors $\psi$ gauged under the fundamental representation of the gauge group and a Higgs-like gauge-singlet $N_f \times N_f$ complex scalar, $H$:
 \begin{align}
\mathcal{L} = \mathcal{L}_K(G_\mu, \lambda_m, \psi, H) + y_H \bar{\psi} H \psi + \text{h.c.} - u_1 \left (\Tr [H H^\dagger]\right)^2 - u_2 \Tr \left [ ( H H^\dagger)^2 \right]  ,
\end{align}
where $\mathcal{L}_K$ summarizes the kinetic terms of the canonically normalized fields.
We consider this theory in the Veneziano limit:
\ea{
X, N_f \to \infty \qquad \text{while}Ê\qquad  N_f/X \equiv x  \quad \text{is kept fixed} \ ,
}
and, in order to define a finite theory in this limit, we rescale the couplings accordingly:
\begin{equation}
a_g = \frac{g^2X}{(4 \pi)^2} \ ,~\ a_H = \frac{y_H^2X}{(4 \pi)^2}\ ,~ z_1 = \frac{u_1N_f^2}{(4 \pi)^2}\  , ~z_2 = \frac{u_2N_f}{(4 \pi)^2} \ .
\end{equation}
We have shown in \cite{Antipin:2011aa} that the model features Banks-Zaks fixed points perturbative in $\epsilon$ with $x = \frac{9}{2}(1-\epsilon)$ and $\frac{9}{2}$ the value of $x$ above which asymptotic freedom is lost. We have also shown that depending on the boundary conditions in the UV for the bare scalar couplings the model can trigger spontaneous symmetry breaking via the CW mechanism. We refer to our original work for the details of the computation. Here we simply add the further constraint of the one-loop Veltman condition to examine the fate of the PNC improved model. 

Consider the classical background field that can break chiral symmetry to the diagonal subgroup\footnote{This choice is the only one giving the global minimum of the tree level potential in the Veneziano limit \cite{Paterson:1980fc, Gildener:1976ih, Antipin:2011aa}.} 
$H_c= \frac{\phi_c}{\sqrt{2N_f}} \mathbf{1}$. On this background the tree-level mass matrix is:
\be
\label{backmasses}\frac{M^2(\phi_c)}{\phi_c^2} =\frac{(4 \pi)^2}{N_f^2}\times \text{diag}\left \{3(z_1+z_2) ,\ z_1+z_2,\ z_1+3z_2, \ \frac{x}{2}a_H \right \} \ ,
\ee
for the masses of respectively the background scalar field, the $N_f^2$ Goldstone bosons, the $N_f^2-1$ scalar bosons orthogonal to the background direction, and the $N_f$ Dirac fermions.
Along the background (and soon to be flat) direction the potential collapses to 
\ea{
V_0=  \frac{(4 \pi)^2}{N_f^2}(z_1+z_2) \phi_c^4 \ ,
\label{effpot}
}
and therefore the flatness condition for the renormalised couplings reads: $z_1(\mu_0)+z_2(\mu_0)=0$. 

Now we impose the Veltman condition Eq.~\eqref{Veltman} leading to
\be
z_1(\mu_0)+2z_2(\mu_0)-a_H(\mu_0)=z_2(\mu_0)-a_H(\mu_0)=0 \ ,
\ee
where in the second equality we applied the renormalisation condition. 
The one-loop CW mass of the scalar along the flat direction and in the Veneziano limit reads:
\begin{equation}
m_l^2=\frac{32\pi^2 \phi_c^2}{N_f^2}\left [(z_1+z_2)^2+(z_1+3z_2)^2 - x a_H^2   \right] \stackrel{z_1=-z_2}{=}\frac{32\pi^2 \phi_c^2}{N_f^2} \left [4z_2^2 - x a_H^2   \right]\stackrel{z_2=a_H}{=}\frac{32\pi^2 \phi_c^2}{N_f^2}z_2^2 \left [4 - x\right ].
\label{Dilatonmass}
\end{equation}
In the second equality we imposed the flatness condition. Here, as noted in \cite{Antipin:2011aa}, one observes that it is possible to enact a CW mechanism even for small $\epsilon$ which selects a region of spontaneous chiral (and henceforth conformal) symmetry breaking near the IR fixed point, also controlled by $\epsilon$. However, as soon as we impose the full PNC conditions (i.e. also the Veltman) we arrive at the last expression on the right hand side.  This condition states that the PNC extremum is a local maximum along the flat direction for the theory controlled by the nearby Banks-Zaks IR fixed point (i.e. $\epsilon\ll 1$, i.e. $x$ very close to 4.5). We can therefore conclude that either the theory remains conformal in the IR or a non-perturbative stable ground state is reached away from the CW perturbative regime. In any case the PNC improved model does not lead automatically to a perturbative CW mechanism.

\end{document}